\documentclass[aps,pra,showpacs,twocolumn,groupedaddress]{revtex4}

\usepackage{graphicx,bbm,amsthm, amsmath, amssymb}
\usepackage{color}

\def\ket#1{|#1\rangle}

\def\pr#1{\tilde{#1}}
\def\braket#1#2{\langle #1 | #2 \rangle}

\def\tit#1{}
\def\etal#1{ {\em et.al.}}
\def\bfth{\boldsymbol{\theta}}

\def\rmi{{\rm i}}

\def\calc{\mathcal{C}}
\def\rmd{\rm d}

\begin{document}

\title{Quantum circuit for three-qubit random states}

\author{Olivier Giraud$^{1,2}$, Marko \v Znidari\v c$^3$ and Bertrand Georgeot$^{1,2}$}

\affiliation{
${}^1$ Universit\'e de Toulouse; UPS; 
Laboratoire de Physique Th\'eorique (IRSAMC); F-31062 Toulouse, France\\
${}^2$ CNRS; LPT (IRSAMC); F-31062 Toulouse, France\\
${}^3$ Department of Physics, Faculty of Mathematics and Physics,\\
University of Ljubljana, SI-1000 Ljubljana, Slovenia
}
 
\date{March 24, 2009}

\begin{abstract}
We explicitly 
construct a quantum circuit which exactly generates random three-qubit 
states. The optimal circuit consists of three CNOT gates and fifteen single 
qubit elementary rotations, parametrized by fourteen independent angles. 
The explicit distribution of these angles is derived, showing that 
the joint distribution is a product of independent distributions 
of individual angles apart from four angles.  
\end{abstract}


\maketitle

\section{Introduction}

Quantum information science (see e.g.~\cite{nielsen} and references therein)
has received an increased attention in recent years due to the understanding 
that it enables to perform procedures not possible by purely classical 
resources. Experimental techniques to manipulate 
increasingly complex quantum systems are also rapidly progressing. 
One of the central issues is on the one hand to control and manipulate delicate 
complex quantum states in an efficient manner, but on the other hand
 at the same time to prevent all uncontrollable influences from the environment.
In order to tackle such problems, one has to understand the structure 
and properties of quantum states. This can be done
 either through studies of particular states 
in a particular setting, or through focusing on the properties of the most
generic states. 

Random quantum states, that is states distributed according to the unitarily 
invariant Fubini-Study measure, are good candidates for describing generic 
states. Indeed, they are typical in the sense that statistical properties of 
states from a given Hilbert space are well described by those of random 
quantum states. Also, they describe eigenstates of sufficiently complex 
quantum systems~\cite{Haake} as well as time evolved states after sufficiently 
long evolution. Not least, because random quantum states possess a large amount
 of entanglement they are useful in certain quantum information processes like 
quantum dense coding and remote state preparation~\cite{Harrow:04,rsp}. They 
can be used to produce random unitaries needed in noise 
estimation~\cite{Emerson} and twirling operations~\cite{toth}.  
In addition, as random states
are closely connected to the unitarily invariant Haar 
measure of unitary matrices, the unitary invariance makes theoretical
treatment of such states simpler.

Producing such states therefore enables to make available a useful
quantum resource, and in addition to span the space of quantum states
in a well-defined sense.  Therefore several works have recently explored 
different procedures to achieve this goal.
It is known that generating random states distributed according to the exact invariant 
measure requires a number of gates exponential in the number of
qubits. A more efficient but approximate way to generate
random states uses pseudo-random quantum circuits in which gates are randomly
drawn from a universal set of gates. As the number of applied gates
increases the resulting measure gets increasingly close to the
asymptotic invariant measure~\cite{Eme03}. Some bipartite properties
of random states can be
reproduced in a number of steps that is smaller than exponential in
the number of qubits.  Polynomial convergence bounds have been
derived analytically for bipartite
entanglement~\cite{OliDahPle07,HarLow08,exact} 
for a number of pseudo-random protocols. On the numerical side, 
different properties of circuits generating random states have been studied 
\cite{numerical}.
In order to quantify how well a given pseudo-random scheme reproduces 
the unitarily invariant distribution, one can study averages of low-order 
polynomials in matrix elements~\cite{converg}. In particular, one can
define a state or a unitary 
$k$-design, for which moments up to order $k$ agree with the Haar 
distribution~\cite{2design,AmbEme07}. Although exact state
$k$-designs can be built for 
all $k$ (see references in~\cite{AmbEme07})
they are in general inefficient. In contrast, efficient approximate 
$k$-designs can be constructed for arbitrary $k$ (for the specific case 
of 2-design see~\cite{HarLow08}).

The pseudo-random circuit approach can yield only
pseudo-random states, which do not reproduce exactly
the unitarily invariant distribution.
The method has been shown to be useful for large number of qubits,
where exact methods are clearly inefficient. However, for systems with
few qubits, the question of asymptotic complexity is not relevant.
It is thus of interest to study specifically these systems and 
to find the most 
efficient way -- in terms of number of gates -- 
to generate random states distributed according to the unitarily invariant
measure. This question is not just of academic interest
since, as mentioned, 
few-qubit random unitaries are needed for e.g. noise estimation or
twirling operations. Optimal circuits for small number of
qubits could also be used as a basic building block of pseudo-random
circuits for larger number qubits, which might lead to faster
convergence.  In addition, systems of few qubits are becoming
available experimentally, and it is important 
to propose algorithms that could be implemented on such small
quantum processors, and which use as
little quantum gates as possible. Indeed, quantum gates, and
especially two-qubit gates, are a scarce resource in real systems
which should be carefully optimized.

In this paper we therefore follow a different strategy from the more
 generally adopted approach of using
pseudo-random circuits to generate pseudo-random
states, and try and construct exact algorithms generating random
states for systems of three qubits. In the language of $k$-designs
such algorithms are exact $\infty$-designs.  We present a circuit composed of
one-qubit and two-qubit gates
which produces exact random states 
in an optimal way, in the sense of using the smallest
possible number of CNOT gates.  The circuit uses in total $3$ CNOT
gates and $15$ one-qubit elementary rotations. 
Our procedure uses results recently
 obtained~\cite{ZniGirGeo08} which described optimal procedures 
to transform a three-qubit state into another.  
Our circuit needs $14$ random numbers which should be classically
 drawn and used as parameters for performing the one-qubit gates.  The
probability distribution of these parameters is derived,
 showing that it factorizes into a product of 10 independent 
distributions of one parameter and a joint distribution of the 4 remaining 
ones, each of these distributions
being explicitly given.
Since we had to devise specific methods to compute these
distributions, we explain the derivation in some details, as these methods
can be useful in other contexts.

After presenting the main idea of the calculation in Section 
\ref{metrictensor}, we start by treating the simple case of two-qubit states
in Section \ref{2qubits}. We then turn to the three-qubit case and first 
show factorization of the probability distribution for a certain subset
of the parameters (Section \ref{7-14}), the remaining parameters being 
treated in Section 
\ref{1-6}. The full probability distribution for three qubits is summarized
in Section \ref{explicit_calc}.

\section{The quantum circuit}
\label{metrictensor}
Formally, a quantum state $\ket{\psi}$ 
can be considered as an element of the complex projective space  
$\mathbbm{C}\mathbf{P}^{N-1}$, with $N=2^n$ the Hilbert 
space dimension for $n$ qubits~\cite{book}. 
The natural Riemannian metric on $\mathbbm{C}\mathbf{P}^{N-1}$ is the 
Fubini-Study metric, induced by the unitarily invariant Haar measure on 
$U(N)$.
It is the only metric invariant under unitary transformations. 
To parametrize $\mathbbm{C}\mathbf{P}^{N-1}$
 one needs $2N-1$ independent real parameters. 
Such parametrizations are
well-known, for instance using Hurwitz parametrization of $U(N)$ \cite{Hur}. 
However, they do not easily translate into one and two-qubit operations,
as desired in quantum information. 
In Ref.~\cite{ZniGirGeo08}, optimal quantum circuits transforming 
the three-qubit state
$\ket{000}$ into an arbitrary quantum state were discussed.
In the case of three qubits, a generic
state can be parametrized up to a global phase by 14 parameters. 
The quantum circuit requiring the smallest amount of CNOT gates
has three CNOTs and 15 one-qubit gates depending on 14 independent rotation 
angles. From \cite{ZniGirGeo08} it is possible (see Appendix) to extract
the circuit depicted in Fig.~\ref{circuit}, expressed as a series 
of CNOT gates and single qubit rotations, where Z-rotation is 
$Z_{\theta}=\exp{(-{\rm i} \sigma_{\rm z} \theta)}$ and 
Y-rotation is 
$Y_{\theta}=\exp{(-{\rm i} \sigma_{\rm y} \theta)}$ with 
$\sigma_{\rm y,z}$ the Pauli matrices.
\begin{figure*}
\begin{center}
\includegraphics[width=.98\linewidth]{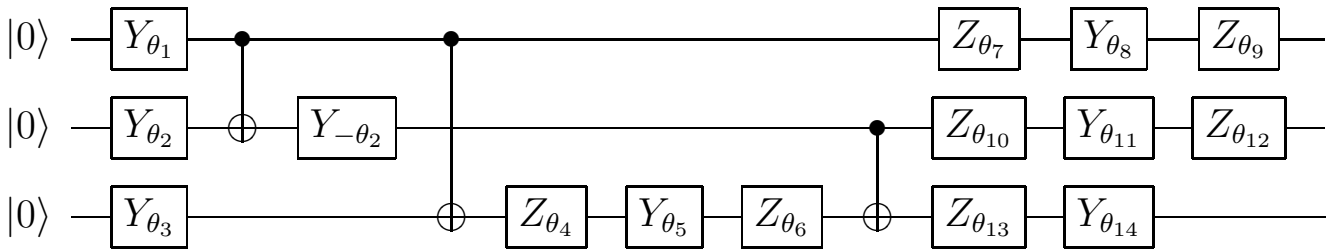}
\end{center}
\caption{Circuit $\calc$ for three-qubit random state generation.}
\label{circuit}
\end{figure*}
The circuit allows to go from $\ket{000}$ 
to any quantum state (up to an irrelevant global phase). It therefore 
provides a parametrization of a quantum state $\ket{\psi}$ by 
angles $\theta_1,\ldots,\theta_{14}$. 

In order to generate random vectors distributed according to the 
Fubini-Study measure, it would of course be 
possible to use e.~g.~Hurwitz parametrization to generate classically a
random state, and then use the procedure described in \cite{ZniGirGeo08}
to find out the consecutive steps that allow to construct this particular 
vector from $\ket{000}$. However this procedure requires application 
of a specific algorithm for each realization of the random vector.
Instead, our aim here is to directly find the distribution of the $\theta_i$ 
such that the resulting $\ket{\psi}$ is distributed according to the 
Fubini-Study measure. 
This is equivalent to calculating the invariant measure associated 
with the parametrization provided by Fig.~\ref{circuit}
in terms of the angles $\theta_1,\ldots,\theta_{14}$.
Geometrically, the Fubini-Study distance $D_{\rm FS}$ is the angle between 
two normalized states, $\cos{(D_{\rm FS})}=|\braket{\psi}{\phi}|$. The 
metric induced by this distance is obtained by taking 
$\ket{\phi}=\ket{\psi}+\ket{\rmd\psi}$, getting
\begin{equation}
\rmd s^2=\frac{\langle\psi,\psi\rangle\langle \rmd\psi,\rmd\psi\rangle-
\langle\psi,\rmd\psi\rangle\langle \rmd\psi,\psi\rangle}{\langle\psi,\psi\rangle^2}
\label{ds2}
\end{equation}
where $\langle,\rangle$ is the usual Hermitian scalar product on 
$\mathbb{C}^N$ \cite{Arn78}. If a state $\ket{\psi}$ is parametrized
by some parameters
 $\theta_1,\theta_2,\ldots$ then the Riemannian metric tensor 
$g_{ij}$ is such that $\rmd s^2=\sum g_{ij}\rmd\theta_i\rmd\theta_j$ and the
volume form at each point of the coordinate patch, directly giving the 
invariant measure, is then given by 
$\rmd v=\sqrt{\det(g)}\prod \rmd\theta_i$.
Thus the joint distribution $P(\bfth)$ of the $\theta_i$ is simply
obtained by calculating the determinant of the metric tensor given by
\eqref{ds2} with the parametrization $\ket{\psi}=\ket{\psi(\bfth)}$,
$\bfth=(\theta_1,\ldots,\theta_{14})$. 
Unfortunately the calculation of such a $14\times 14$ determinant for 
$n=3$ qubits is intractable and one has to resort to other means. Let us 
first consider the easier case of $n=2$ qubits, where by contrast the
calculation can be performed directly.
\begin{figure}[th!]
\begin{center}
\includegraphics[width=.98\linewidth]{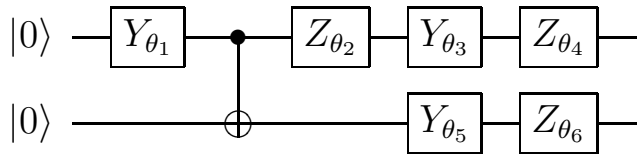}
\end{center}
\caption{Circuit for two-qubit random state generation. }
\label{fig:2qcircuit}
\end{figure}

\section{A simple example: The two-qubit case}
\label{2qubits}
A normalized random 2-qubit state $\ket{\psi}$ depends, up to a global phase,
on 6 independent real parameters. 
A circuit producing $\ket{\psi}$ from an initial state
 $\ket{00}$ is depicted in Fig.~\ref{fig:2qcircuit}. 
One can easily calculate the parametrization of the final state 
$\ket{\psi(\theta_1,\ldots,\theta_6)}$ in terms of all six angles,
thus directly obtaining the metric tensor $g_{ij}$ from (\ref{ds2}). Square 
root of the determinant of $g$ then gives an unnormalized probability 
distribution of the angles as
\begin{equation}
P(\bfth)=|\cos^2{2\theta_1}\sin{2\theta_1}\sin{2\theta_3}
\sin{2\theta_5}|
\label{eq:2qP}
\end{equation}
(see also ~\cite{Bengtsson}). Several observations can be made about this
distribution. First, the three rotations applied on the first qubit 
(top wire in Fig.~\ref{fig:2qcircuit}) after the CNOT gate represent a random 
SU(2) rotation, for which the Y-rotation angle is distributed as 
$P(\theta_3) \sim |\sin{2\theta_3}|$ and the Z-rotation angles are
uniformly distributed \cite{Hur}. Secondly, angle $\theta_1$ 
gives the eigenvalue of the reduced density matrix, 
$\lambda=\sin^2\theta_1$, 
for which the distribution is well known, see, {\em e.g.},
~\cite{ZycSom01}. The third observation is that, remarkably, 
the joint distribution (\ref{eq:2qP}) of 
all 6 angles factorizes into 6 independent one-angle distributions.

\section{Factorization of the three-qubit distribution for angles $\theta_7$
to $\theta_{14}$}
\label{7-14}
Let us now turn to our main issue, which is the distribution of
angles in the three-qubit case.
In order to have an indication whether the distribution of an angle $\theta_i$
factorizes, 
we numerically computed the determinant det$(g)$ of the metric tensor  
as a function of $\theta_i$ with the other angles fixed. We also 
numerically computed the marginal distribution of $\theta_i$ by using 
the procedure given in the 
appendix to find the angles corresponding to a sample of uniformly
distributed random vectors. If the distribution for a given angle 
$\theta_i$ factorizes, these two numerically computed functions 
should match (up
to a constant factor). This is what we observed for all angles but four of
them (angles $\theta_3$ to $\theta_6$).

In order to turn this numerical observation into a rigorous proof,
we are going to show in this section that the distributions for
angles $\theta_{7}$ to $\theta_{14}$ indeed factorize. 
In the next section we will complete the proof by dealing with the cases
$\theta_1$ to $\theta_6$. The explicit analytical expression of the 
probability distribution for individual angles will be given in 
Section \ref{explicit_calc}.
 
Let us denote by $\mathcal{C}$ the circuit of Fig.~\ref{circuit}
 and by $C(\bfth)$ the unitary operator corresponding to it, so that 
$\ket{\psi(\bfth)}=C(\bfth)\ket{000}$. Because circuits $\mathcal{C}$ 
span the whole space of 3-qubit states, any 
unitary 3-qubit transformation $V$ maps parameters $\bfth$
to new parameters $\pr{\bfth}$ such that 
$V\ket{\psi(\bfth)}=\ket{\psi(\pr{\bfth})}$. 
We denote by $\pr{\calc}$ the circuit parametrized by angles $\pr{\bfth}$ 
corresponding to performing $\calc$ followed by $V$. 
It is associated with the unitary operator $C(\pr{\bfth})$ such that
$C(\pr{\bfth})\ket{000}=V\, C(\bfth)\ket{000}$. 
Unitary invariance of the measure implies for
$P(\bfth)$ that 
\begin{equation}
\label{eq:invariance}
P(\bfth)=P(\pr{\bfth})|\mathcal{J}|,
\end{equation}
with $\mathcal{J}$ the Jacobian of the transformation
$\bfth\mapsto\pr{\bfth}$
and $|.|$ denotes the determinant.  
Note that Eq.~(\ref{eq:invariance}) is not a simple change of variables,
as the same function $P$ appears on both sides of the equation. 
The Jacobian matrix $\mathcal{J}$ for transformation $V$ from angles $\bfth$
to $\pr{\bfth}$, $V\ket{\psi(\bfth)}=\ket{\psi(\pr{\bfth})}$, tells how 
much do the 
 angles $\pr{\bfth}$ of $\ket{\psi(\pr{\bfth})}$ change if we vary angles 
$\bfth$ 
in $\ket{\psi(\bfth)}$ keeping transformation matrix $V$ fixed. Choosing $V$
that sets some angles $\theta_j$ in circuit $\pr{\calc}$ to a fixed value,
say zero, and at the same time showing that $|\mathcal{J}|$ depends only on
these angles $\theta_j$, would prove factorization of $P(\bfth)$ with
respect to angles $\theta_j$ through Eq.~\ref{eq:invariance}.

\subsection{Gates 7-12 and 14}
\label{gate14}
The simplest case is that of gates at the end of the circuit 
$\calc$ of Fig.~\ref{circuit}, {\em e.g.}, 
gate $Y_{\theta_{14}}$. For $V$ we take Y-rotation by angle $-u$ on the third qubit, $V=Y_{-u}$. It defines a mapping $\bfth\mapsto\pr{\bfth}$
such that $\pr{\theta}_i=\theta_i$ for $i\leq 13$ and
$\pr{\theta}_{14}=\theta_{14}-u$. Matrix elements of the Jacobian,
i.e. partial derivatives $\mathcal{J}_{jk}=\partial \pr{\theta}_j/\partial
\theta_k$, are equal to the Kronecker symbol $\delta_{jk}$. 
The Jacobian is equal to an identity matrix and its determinant is one. Equation~\eqref{eq:invariance} taken at
$u=\theta_{14}$ then gives
$P(\theta_1,\ldots,\theta_{13},0)=P(\theta_1,\ldots,\theta_{13},\theta_{14})$,
from which one concludes that the distribution for $\theta_{14}$ factorizes
 and is in fact uniform (unless noted otherwise $P$'s are not normalized).
 The same argument holds for the two other rotations by angles $\theta_{12}$
 and $\theta_{9}$ applied at the end of each qubit wire. 

Proceeding to angle $\theta_8$ one could use 
$V=Y_{-u_8}Z_{-u_9}$ applied on the first qubit and show that the Jacobian
depends only on $\theta_8$ and $\theta_9$, while at $u_8=\theta_8$ and
$u_9=\theta_9$ one gets $\pr{\theta_8}=\pr{\theta_9}=0$, from which
factorization of $\theta_8$ would follow from Eq.~\ref{eq:invariance}. 
There is however a simpler way. Observe that the three single-qubit gates 
with angles $\theta_7,\theta_8$ and $\theta_9$ on the first qubit 
span the whole SU(2) group. Therefore, for any one-qubit unitary $V$, gates
$V Z_{\theta_9}Y_{\theta_8}Z_{\theta_7}$ can be rewritten as 
$Z_{\pr{\theta}_9}Y_{\pr{\theta}_8}Z_{\pr{\theta}_7}$, 
without affecting other $\theta$'s. The distribution of these three angles 
must therefore be the same as the distribution of corresponding SU(2) 
parameters. Note that the same argument can be applied in the 
 2-qubit case of Section \ref{2qubits}. As a consequence, the distribution
of angles for gates $Z-Y-Z$ at the end of the circuit
should be the same in both cases, 
that is the distribution of $\theta_7$ is uniform while that of $\theta_8$ 
is proportional to $|\sin{2\theta_8}|$. Similarly, one 
can show that the distribution for the angles $\theta_{10}$ to $\theta_{12}$
 is the same as for angles $\theta_7$ to $\theta_9$.

\subsection{Gate 13}
As opposed to gates 10-12, for gate 13 we can not use the analogy 
with the 2-qubit circuit (Fig.\ref{fig:2qcircuit}) because the two gates 
13 and 14 on the third qubit do not span the whole SU(2) group. Therefore
a different argument should be used.
In what follows we show that the joint distribution for angles $\theta_{13}$
and $\theta_{14}$ can be factorized out of the full distribution. 
Since it has been shown in Subsection \ref{gate14} that angle $\theta_{14}$
 factorizes, this will prove that the distribution for $\theta_{13}$ 
also factorizes.

Using $V=Z_{-u_{13}}Y_{-u_{14}}$ on the third qubit we can 
set $\pr{\theta}_{13}$ and $\pr{\theta}_{14}$ to zero with the choice
$u_{13}=\theta_{13}$ and $u_{14}=\theta_{14}$. Our goal is to show that $|\mathcal{J}|$ depends only on $\theta_{13}$ and $\theta_{14}$. We can formally consider each angle $\pr{\theta}_i$ as being a 
function $\pr{\theta}_i(\bfth;u_{13},u_{14})$ of the
 initial $\bfth$ as well as of the parameters $u_{13,14}$ through 
$C(\pr{\bfth})\ket{000}=VC(\bfth)\ket{000}$. To calculate matrix 
elements of $\mathcal{J}$ for our choice of 
$V$ evaluated at $u_{13}=\theta_{13}$ and $u_{14}=\theta_{14}$,
we must obtain the first-order expansion in $\epsilon$ of the quantities
\begin{equation}
\label{limlim}
\pr{\theta}_j(\theta_1,\ldots,\theta_{k-1},\theta_k+\epsilon,\theta_{k+1},\ldots,\theta_{14};\theta_{13},\theta_{14}).
\end{equation}
Some angles $\pr{\theta}_j$ are very simple. 
We immediately see that when varying
angles $\theta_{k \leq 12}$, that is taking $k \leq 12$ in Eq.\ref{limlim},
angles $\theta_{j \leq 12}$ do not change (i.e 
$\pr{\theta}_j=\theta_j$). The corresponding $12 \times
12$-dimensional subblock in $\mathcal{J}$ is therefore equal to an identity
matrix. Similarly, taking $k=13$ in Eq.\ref{limlim} we see that 
$\theta_{j \leq 13}$ do
not change. The corresponding column in $\mathcal{J}$ is therefore 
zero apart from $1$
on the diagonal. The Jacobian thus has a block structure of the form
\begin{equation}
|\mathcal{J}|=\begin{vmatrix}
\mathbbm{1} & B \\
0 & A
\end{vmatrix} =|A|,
\label{eq:Jblock}
\end{equation}  
where $\mathbbm{1}$ is a $13\times 13$-dimensional identity matrix and 
$A$ is a $1 \times 1$-dimensional block with partial derivative
 $\partial \pr{\theta}_{14}/\partial \theta_{14}$. The angle $\pr{\theta}_{14}$ given by Eq.(\ref{limlim}) is obtained
by varying angle $\theta_{14}$ by $\epsilon$. 
The condition that $C(\pr{\bfth})\ket{000}=VC(\bfth)\ket{000}$
is $Y_{\pr{\theta}_{14}}Z_{\pr{\theta}_{13}}\ket{\pr{x}}
=Z_{-\theta_{13}}Y_{\epsilon}Z_{\theta_{13}}\ket{x}$, 
where $\ket{x}={\rm CNOT}_{23}\ket{\phi}$ is a state after 
the third CNOT acts on $\ket{\phi}$ (counting from the left 
in Fig.~\ref{circuit}), explicitly given by
\begin{equation}
\label{finalform}
\ket{\phi}=\cos\theta_1\ket{00\alpha}+\sin\theta_1\ket{1}
\left(\sin 2\theta_2\ket{0}+\cos 2\theta_2\ket{1}\right)\ket{\beta}
\end{equation}
with
\begin{eqnarray}
\label{alphabeta}
\ket{\alpha}&=&\left(\cos\theta_3\cos\theta_5
-e^{2 \rmi \theta_4}\sin\theta_3\sin\theta_5\right)\ket{0}\\
&+&e^{2 \rmi \theta_6}
\left(\cos\theta_3\sin\theta_5
+e^{2 \rmi \theta_4}\sin\theta_3\cos\theta_5\right)\ket{1}\nonumber\\
\ket{\beta}&=&\left(\sin\theta_3\cos\theta_5
-e^{2 \rmi \theta_4}\cos\theta_3\sin\theta_5\right)\ket{0}\\
&+&e^{2 \rmi \theta_6}
\left(e^{2 \rmi \theta_4}\cos\theta_3\cos\theta_5
+\sin\theta_3\sin\theta_5\right)\ket{1}.\nonumber
\end{eqnarray} 
$\ket{x}$ is therefore determined by $\alpha,\beta$ and $\bar{\beta}$, 
where $\ket{\bar{\beta}}=\sigma_{\rm x}\ket{\beta}$. 
Similarly, $\ket{\pr{x}}$ is determined by $\pr{\alpha},\pr{\beta}$ and 
$\pr{\bar{\beta}}$. Projecting these conditions for $\ket{\beta}$
and $\ket{\bar{\beta}}$ on the computational basis $\{\ket{0},\ket{1}\}$, 
eliminating unwanted variables, we get the following two equations:
\begin{equation}
\frac{\cos{\epsilon}-\kappa\sin{\epsilon}\, 
{\rm e}^{2\rmi\theta_{13}}}{\cos{\epsilon}+\kappa\sin{\epsilon}\, 
{\rm e}^{-2\rmi\theta_{13}}}
=\frac{\cos\pr{\theta}_{14}{\rm e}^{-\rmi\pr{\theta}_{13}}
-\kappa\sin\pr{\theta}_{14}{\rm e}^{\rmi\pr{\theta}_{13}}}{\cos\pr{\theta}_{14}{\rm e}^{\rmi\pr{\theta}_{13}}+\kappa\sin\pr{\theta}_{14}{\rm e}^{-\rmi\pr{\theta}_{13}}},
\end{equation}
one with $\kappa=1$, one with $\kappa=-1$.
Expanding these equations to first order in $\epsilon$ yields
$\pr{\theta}_{14}=\epsilon \cos 2\theta_{13}+o(\epsilon^2)$ and $\pr{\theta}_{13}=0+o(\epsilon^2)$. The derivative
 $\partial\pr{\theta}_{14}/\partial\theta_{14}$ is therefore equal to $\cos 2\theta_{13}$, which completes the proof. 
Incidentally, we also see that the distribution of $\theta_{13}$ is 
proportional to $|\cos{2\theta_{13}}|$.

\section{Joint three-qubit probability distribution for angles $\theta_1$ to
  $\theta_6$}
\label{1-6}
In the preceding section, we have shown that the distribution
for angles $\theta_7$ to $\theta_{14}$ factorizes. As was mentioned,
numerical observations indicated us that the distribution for angles
$\theta_1$ and $\theta_2$ should also factorize, but that it is not
the case for the joint distribution of $\theta_3, \ldots,\theta_6$.

As we were not able to directly prove by the same methods as above
that the distributions for $\theta_1$ and $\theta_2$ factorize,
we use a different strategy. Namely, we first assume that this 
factorization is true, 
then we compute the distributions under this assumption, and the knowledge
of the answer allows us to prove a posteriori that it is indeed
the correct probability distribution.

If the factorization holds,
the distribution for $\theta_1$ and $\theta_2$ is easily calculated
from the matrix $g$ using symbolic manipulation software,
by replacing angles $\theta_j$, $j\geq 3$, in $g$ by suitably chosen 
simple values, so that the $14\times 14$ determinant giving the volume 
form can now be handled.
This yields, up to a normalization constant,
\begin{eqnarray}
\label{p1}
P_1(\theta_1)&=&\cos^5\theta_1\sin^9\theta_1\\
P_2(\theta_2)&=&\cos^5 2\theta_2\sin^3 2\theta_2.
\label{p2}
\end{eqnarray}

The joint distribution of $\theta_3, \ldots,\theta_6$ can not be further factorized,
and requires heavy calculations. Indeed,
even replacing all angles but $\theta_3,...,\theta_6$ by numerical
values the determinant $\det(g)$ of the metric tensor given by \eqref{ds2}
still depends on 4 variables, which is too much for it to be evaluated
by standard software. We thus proceed as follows. First one can
show that $\det(g)$ can be put under the form
\begin{eqnarray}
\label{detg}
\det(g)=\sum_{p=-10}^{10}\sum_{q=-6}^{6}\sum_{r=-8}^{8}\sum_{s=-6}^{6}&
a_{pqrs}\cos(2p\theta_3+2q\theta_4\nonumber\\
&+2r\theta_5+2s\theta_6),
\end{eqnarray}
with the sums running over all $q,r$ but only even values of $p$ and $s$. 
Because of the parity of $\cos$, there are $M=8509$ independent coefficients
$a_{pqrs}$. Evaluating numerically the determinant at $M$ random values
of the angles one gets an $M\times M$ linear system that can be solved
numerically. If the values of the coefficients of the matrix $g_{ij}$ 
are multiplied by a factor 4, then one is ensured (from inspection of 
$\det(g)$)
that the $a_{pqrs}$ 
are rationals of the form $k/2^9$, $k\in\mathbb{Z}$. This allows to
deduce their exact value from the numerical result.
We are left with 6998 nonzero terms in $\det(g)$, and terms with
odd $q$ or $r$ do not exist. We then suppose that $\sqrt{\det(g)}$ can 
be expanded as 
\begin{eqnarray}
\label{sqrtdetg}
\sqrt{\det(g)}=\hspace{6cm}\\
\sum_{p=-5}^{5}\sum_{q=-3}^{3}\sum_{r=-4}^{4}\sum_{s=-3}^{3}
b_{pqrs}e^{\mathrm{i}(2p\theta_3+2q\theta_4+2r\theta_5+2s\theta_6)}.\nonumber
\end{eqnarray}
This assumption is validated a posteriori, since a solution
of the form \eqref{sqrtdetg} can indeed be found.
There are 4851 coefficients $b_{pqrs}$, which can be obtained by
identifying term by term coefficients in the expansion of 
$(\sqrt{\det(g)})^2$ and $\det(g)$. We have to solve a system of 
quadratic equations
\begin{eqnarray}
a_{10,6,8,6}&=&b_{5343}^2\\
a_{10,6,8,5}&=&b_{5342}b_{5343}+b_{5343}b_{5342}\nonumber\\
a_{10,6,8,4}&=&b_{5341}b_{5343}+b_{5342}b_{5342}+b_{5343}b_{5341}\nonumber\\
\ldots&=&\ldots\nonumber
\end{eqnarray}
The first equation is quadratic and fixes an overall sign. 
Equation $k+1$ is linear once the values
obtained from equations $1$ to $k$ are plugged into it.
Starting with the highest-degree term $(p,q,r,s)=(5,3,4,3)$ one 
can thus recursively solve all equations. There are only 1320
non-zero coefficients $b_{pqrs}$. Gathering together terms
$(\pm p,\pm q,\pm r,\pm s)$ one can simplify the sum \eqref{sqrtdetg}
to a sum of 96 terms of the form $c_{pqrs}\cos(p\theta_3)\cos(q\theta_4)
\cos(r\theta_5)\sin(s\theta_6)$. Expanding this expression in powers
of $\cos(2\theta_5)$ and $\sin(2\theta_5)$ and simplifying separately
each coefficient we finally get
\begin{equation}
\label{P3456}
P(\theta_3,\theta_4,\theta_5,\theta_{6})=\sin 2\theta_5
\sin 4\theta_3\sin^2\varphi_1\cos\varphi_2,
\end{equation}
where $\langle\alpha|\bar{\beta}\rangle=\cos\varphi_1$ and
$\langle\beta|\bar{\beta}\rangle=\cos\varphi_2$ with $\ket{\alpha}$, $\ket{\beta}$
given by Eq.~\eqref{alphabeta}. Recall that 
$\ket{\bar{\beta}}$ is the bit-flip transform
of $\ket{\beta}$, $\ket{\bar{\beta}}=\sigma_{\rm x}\ket{\beta}$. Note that 
$\sin2\theta_3=\langle\alpha|\beta\rangle$. 
Angles $\varphi_{1,2}$ can be obtained from
\begin{eqnarray}
\cos^2{\varphi_1}&=&({\rm c}_4{\rm c}_5{\rm c}_6-{\rm s}_4{\rm s}_6)^2+({\rm c}_3{\rm c}_4{\rm s}_6+{\rm c}_3{\rm s}_4{\rm c}_5{\rm c}_6)^2 \nonumber \\
\cos{\varphi_2}&=&-{\rm s}_3{\rm s}_4{\rm s}_6+{\rm c}_6({\rm s}_3{\rm c}_4{\rm c}_5-{\rm c}_3{\rm s}_5),
\label{eq:phi}
\end{eqnarray}
where ${\rm c}_i=\cos{2\theta_i}$ and ${\rm s}_i=\sin{2\theta_i}$.
We do not have a general argument to explain this remarkable 
expression of the distribution in terms of the scalar products
of $\ket{\alpha}$, $\ket{\beta}$ and $\ket{\bar{\beta}}$.

To complete the proof for the joint distribution
$P(\theta_1,\ldots,\theta_6)$ it remains to be checked that the determinant
of the metric tensor $g$ with angles $\theta_7$ to $\theta_{14}$ replaced by
constants is indeed proportional to 
$P_1(\theta_1)P_2(\theta_2)P(\theta_3,\theta_4,\theta_5,\theta_6)$. This
a posteriori verification is easier to handle symbolically than the full
a priori calculation of the $14\times 14$ determinant. Indeed, the
determinant can first be reduced to an $8\times 8$ determinant by
Gauss-Jordan elimination. The remaining determinant can be expanded
as a trigonometric polynomial. Although symbolic manipulation software do
not allow to simplify the coefficients of this polynomial, they are able 
to check that these coefficients match those of the expected distribution.
We proved in that way that the difference between the determinant det$(g)$
and our expression is identically zero. This gives a computer-assisted
but rigorous proof for the distribution of angles $\theta_1$ to $\theta_6$.

\section{Total three-qubit probability distribution function}
\label{explicit_calc}
Gathering together the results of the previous sections we obtain that
the joint distribution $P(\bfth)$ can be factorized as
\begin{equation}
P(\bfth)=|P_1(\theta_1)P_2(\theta_2)P(\theta_3,\theta_4,\theta_5,\theta_{6})
\prod_{i=7}^{14}P_i(\theta_i)|.
\label{eq:P}
\end{equation}
The joint distribution $P(\theta_3,\theta_4,\theta_5,\theta_{6})$
has been derived in the previous section and is given by Eq.~\eqref{P3456}.
The distribution for $\theta_1$ and $\theta_2$ is given by 
Eqs.~\eqref{p1}-\eqref{p2}.
Given the factorization \eqref{eq:P}, it is easy to calculate the remaining
$P_i(\theta_i)$ for each $i=7,\ldots,14$ as was done for $\theta_1$ and 
$\theta_2$ in the previous section: replacing angles
$\theta_j$, $j\neq i$, in $g$ by suitably chosen simple values, 
the $14\times 14$ determinant giving the volume form can be
easily evaluated by standard symbolic manipulation. 
This yields, up to a normalization constant,
\begin{equation}
\prod_{i=7}^{14}P_i(\theta_i)
=\sin 2\theta_8\sin 2\theta_{11}\cos 2\theta_{13}.
\end{equation}

The knowledge of the angle distribution (\ref{eq:P}) allows to
easily generate random three-qubit vectors using the circuit of 
Fig.~\ref{circuit}.
Angles $\theta_1,\theta_2$ and $\theta_7$ to $\theta_{14}$ can be drawn 
classically according to their individual probability distribution. 
Angles $\theta_3\ldots,\theta_6$ can be obtained classically from the joint
distribution \eqref{P3456} by, for instance, Monte-Carlo rejection method 
(that is, drawing angles $\theta_3$ to $\theta_6$ and a parameter 
$x\in[0,p]$ at random, and keeping them if
$P(\theta_3,\theta_4,\theta_5,\theta_{6})<x$). 
Bounding $P(\theta_3,\theta_4,\theta_5,\theta_{6})$ from 
above by $p=0.85$ yields a success rate of about $12\%$.

\section{Conclusion}

In this work, we constructed a quantum circuit for generating
three-qubit states distributed according to the unitarily invariant
measure.  The construction is exact and optimal in the sense of having
the smallest possible number of CNOT gates. The
procedure requires a set of $14$ random numbers classically drawn,
which will be the angles of the one-qubit rotations, and whose
distribution has been explicitly given.
Remarkably, we have shown that the distribution of angles factorizes,
apart from that of four angles. The circuit can be used as
a three-qubit random state generator, thus producing at will
typical states on three qubits.  It could be also used
 as a building block for pseudo-random circuits in order to produce 
pseudo-random quantum states on an arbitrary number of qubits. At
last, it gives an example of a quantum algorithm producing interesting 
results which could be implemented on a few-qubit platform, using
only $18$ quantum gates, of which $15$ are one-qubit elementary rotations 
much less demanding experimentally.

\vspace{2cm}
We thank the French ANR (project INFOSYSQQ) and the IST-FET 
program of the EC(project EUROSQIP) for funding.
M\v Z would like to acknowledge support by Slovenian Research Agency, 
grant J1-7437, and hospitality of Laboratoire de Physique Th\' eorique, 
Toulouse, where this work has been started.

\appendix*

\section{The parametrization corresponding to the three-qubit circuit}

In this Appendix, we explain how to obtain the angles $\theta_i$ of
the circuit (Fig.~\ref{circuit}) for a given $\ket{\psi}$, based on
the  discussion in~\cite{ZniGirGeo08}. This justifies the use of these
angles as a parametrization of the quantum states.  We start from a
state $\ket{\psi}$, and transform it by the inverse of the different gates of
Fig.~\ref{circuit} to end up with $\ket{000}$, specifying how the
angles $\theta_i$ are obtained in turn.  More details can be found 
in~\cite{ZniGirGeo08}.
A generic three-qubit state $\ket{\psi}$ can be written in a canonical
form as a sum of two (not normalized) product terms~\cite{AciAndJan01},
\begin{equation}
\ket{\psi}=\ket{\omega_1 \omega_2 
\omega_3}+\ket{\omega_{1}^{\perp}}\ket{\xi}_{23},
\label{eq:canon}
\end{equation}
where $\ket{\omega_i}$ are one-qubit states,  $\ket{\omega_{1}^{\perp}}$ 
is a one-qubit state
orthogonal to $\ket{\omega_1}$ and
$\ket{\xi}_{23}$ is a two-qubit state of the second and third qubits. 
The angle $\theta_9$ is chosen such that the Z-rotation of angle
$-\theta_9$ eliminates a relative phase
between the coefficients of the expansion of $\ket{\omega_1}$ into 
$\ket{0}$ and $\ket{1}$.
(Note that because we are using the circuit in the reverse direction the
angles of rotations have opposite signs). 
A subsequent Y-rotation with angle $-\theta_8$ results in the
transformation $\ket{\omega_1} \rightarrow \ket{0}$ (up to a global
phase). Similarly, rotations of angles $-\theta_{12}$ and $-\theta_{11}$ 
rotate
$\ket{\omega_2}$ into $\ket{0}$. 
After applying rotations of angles $-\theta_8$, $-\theta_9$,
$-\theta_{11}$ and $-\theta_{12}$ the state has become of the form
$\ket{\psi'}=\ket{00\gamma}+\ket{1}(\ket{0\gamma_1}+\ket{1\gamma_2})$ 
(up to normalization). 
Two rotations on the third qubit of angles $-\theta_{13}$ and
$-\theta_{14}$ are now chosen so as to rotate
$\ket{\gamma_1}$ into some new state $\ket{\gamma'}$ while
$\ket{\gamma_2}$ is rotated, up to normalization, into $\sigma_{\rm
  x}\ket{\gamma'}$. It was shown in \cite{ZniGirGeo08} that this can
always be done by 
writing the normalized $\ket{\gamma_{1,2}}$ as
$\ket{\gamma_{1,2}}=\cos{\phi_{1,2}}\ket{0}+{\rm e}^{{\rm i}
  \xi_{1,2}}\sin{\phi_{1,2}}\ket{1}$, and then $\theta_{14}$ is a solution 
of
\begin{equation}
-\tan{(2\theta_{14})}=\frac{\cos{2\phi_1}+\cos{2\phi_2}}{\sin{2\phi_1}\cos{\xi_1}+\sin{2\phi_2}\cos{\xi_2}},
\end{equation}
while $\theta_{13}=-(\delta_1+\delta_2)/4$, where $\delta$'s are
relative phases in $Y_{-\theta_{14}}\ket{\gamma_{1,2}}={\rm
  e}^{{\rm i}\delta_{1,2}} \cos{\kappa}\ket{0}+\sin{\kappa}\ket{1}$. 
Acting with a CNOT$_{23}$ gate on the resulting state one obtains a
quantum state for the three qubits of the form 
$\ket{\psi''}=\ket{00\chi_1}+\ket{1\omega_4\chi_2}$, with $\chi_2=\gamma'$. 
The Z-rotation angle
$-\theta_{10}$ on the second qubit is now determined so as to eliminate 
a relative phase between the expansion coefficients of $\ket{\omega_4}$, 
making them real up to a global phase. On the third qubit we now apply
three rotations of angles $-\theta_4$, $-\theta_5$, and $-\theta_6$ to 
bring 
$\ket{\chi_1}$ to $\ket{\chi'}$ and $\ket{\chi_2}$ into $\sigma_{\rm
  x}\ket{\chi'}$, eliminating also a relative phase. Then a
CNOT$_{13}$ is applied. At this point
(after the second CNOT in Fig.~\ref{circuit}, counting from right, but 
without the $\theta_7$
rotation), the state has become of the form
$\ket{\psi'''}=\cos{\theta_1}\ket{00\omega_6}+{\rm e}^{{\rm
    i}\tau}\sin{\theta_1}\ket{1\omega_5\omega_6}$, where the one-qubit
states $\ket{\omega_5}$ and $\ket{\omega_6}$ are normalized and real. 
With $\theta_7$ we now eliminate the relative phase $\tau$, and
with an Y-rotation of angle $-\theta_3$ the third qubit is brought
to the state $\ket{0}$.  Then the combination of two Y-rotation of
angles $\theta_2$ and $-\theta_2$ with a CNOT$_{12}$ brings the second
qubit to $\ket{0}$, and the last rotation of angle $-\theta_1$ on the
first qubit yields the final state $\ket{000}$.
Note that in the circuit of Fig.~\ref{circuit}
the two Z-rotations of angles $\theta_7$ and $\theta_{10}$
commute with CNOT gates if they act on the control qubit. 
This is the reason why the rotation of angle $\theta_7$ can be applied at 
any point between 
$\theta_1$ and $\theta_8$ and, similarly, $\theta_{10}$ can be applied 
at any point between $\theta_2$ and $\theta_{11}$.

\end{document}